\documentclass[secnumarabic,tightenlines,amssymb, amsmath,nobibnotes,aps,nofootinbib,showpacs]{revtex4}
\usepackage{bm}%
\usepackage{epsfig}%
\usepackage{graphicx}%
\usepackage{amssymb}

\begin{document}

\title{Spin-rotation coupling in compound spin objects}

\author{Gaetano Lambiase$^{a,b,c}$, Giorgio Papini$^{c,d,e}$}
\affiliation{$^a$Dipartimento di Fisica "E.R. Caianiello", Universit\'a di Salerno, 84084 Fisciano (Sa), Italy.}
\affiliation{$^b$INFN, Sezione di Napoli Italy.} 
\affiliation{$^c$International Institute for Advanced Scientific Studies, 89019 Vietri sul Mare (SA), Italy.}
\affiliation{$^d$Department of Physics, University of Regina, Regina, SK, S4S 0A2, Canada.} 
\affiliation{$^e$Prairie Particle Physics Institute, Regina, SK, S4S  0A2, Canada.}

\def\be{\begin{equation}}
\def\ee{\end{equation}}
\def\al{\alpha}
\def\bea{\begin{eqnarray}}
\def\eea{\end{eqnarray}}

\begin{abstract}

We generalize spin-rotation coupling to compound spin systems. In the case of muons bound to nuclei in a storage ring the decay
process acquires a modulation. Typical frequencies for $Z/A\sim 1/2$ are $\sim 3\times 10^6$Hz, a factor $10$ higher than the
modulation observed in $g-2$ experiments.

\end{abstract}


\maketitle

\section{Introduction}

Spin-rotation coupling (SRC) is a necessary ingredient in tests involving inertia and gravity, in physical \cite{cai} and astrophysical
\cite{cai91,papini94} processes, in the generation and control of spin currents and in measurements of the anomalous $g-2$
factor of fermions. It extends our knowledge of rotational inertia to
the quantum level in ways that violate the equivalence principle (EP) \cite{mashhoon2} that is
well-tested experimentally at the classical level.
It can in fact be argued that EP does not apply in the
quantum world. Phase shifts in particle
interferometers \cite{lammer,singh} and particle wave functions certainly depend on the masses of
the particles involved \cite{greenberg,cai91}. Also, EP
does not hold in the context of the causal interpretation of quantum mechanics
 \cite{holl} and several models predict quantum violations of EP
\cite{morgan,peres}, recently in connection with neutrino oscillations
\cite{gasperini,halprin,butler,bozza,adunas}. The Mashhoon term, in particular, yields
different potentials for different particles and different spin states \cite{cai91}
and can not, therefore, be considered universal. Its presence is predicted by all approaches based on the Dirac equation \cite{hehl,cai91,papini94,singh}. The effect
has been measured directly for photons \cite{ashby,mash11} and more recently confirmed to 18\% accuracy for fermions \cite{ni}. The latter result follows
from precise cosmic-spin coupling experiments \cite{heckel} and from nuclear spin gyroscope measurements of the earth rotation \cite{kornack,ni}.

It is useful to examine once more SRC's role in the determination of anomalous magnetic moments.
A relevant point of the $g-2$ experiment is that muons on
equilibrium orbits within a small fraction of the maximum momentum are almost completely
polarized with spin vectors pointing in the direction of motion. As the muons decay,
those electrons projected forward in the muon rest frame can therefore be detected around the ring and
their angular distribution reflects the precession of the muon spin along the
cyclotron orbits.
In the $g-2$ experiment \cite{pap,pap1} it is the Mashhoon
effect \cite{mash} that evidences the $g-2$ term in
 \begin{equation}\label{omegafin}
 \Omega=2\mu B-\omega_2
 =\left(1+\frac{g_{\mu}-2}{2}\right)\frac{eB}{m}-
 \frac{eB}{m}
 = \frac{g_{\mu}-2}{2}\frac{eB}{m}\,,
 \end{equation}
by exactly cancelling, in $2\mu B$, the
much larger Bohr magneton contribution of fermions with no anomalous magnetic
moment. In (\ref{omegafin}), $m$ is the mass of the muon and the result is referred to the rest frame of the muon. The cancellation is made possible by the non-diagonal form of the evolution matrix $M$ of \cite{pap} and is
therefore a direct consequence of the violation of EP.

Recently, the possibility of studying the evolution of heavily charged ions in storage rings \cite{GSI,GSIb,geissel}
raises the question of the role of SRC in rotating compound spin systems \cite{LPSarX,LPSPLB,Pavlichenkov} that have applications
in fields like nuclear
physics, QED, bound state (BS)-QED and stellar nucleosynthesis \cite{takahashi}.

The purpose of this work is to investigate SRC in the semiclassical
system represented by a nucleus plus a decaying charged particle, specifically a negative muon, rotating in a storage ring.
The problem is closely related to the original $g-2$ experimental setup \cite{bailey,farley}.
Though the muon is not just a more massive electron, information about the particle that is dragged along the orbit by the nucleus can be obtained only if the particle itself decays. We apply to this problem some
of the results obtained in \cite{LPSPLB}.

\section{Spin precession for bound muons}
\label{spinprec}

We refer our calculation to a frame which rotates about the $x_3$-axis
in the clockwise direction of an ion in a Storage Ring, with the $x_2$-axis
tangent to the ion orbit in the direction of its momentum and write ${\bf B}=B{\bf \hat u}_3$. The main elements
of the calculation are
${\bm\omega}_{g_{\mu}}$ that represents the coupling of the magnetic moment of the muon to the
magnetic field in the Experimental Storage Ring, ${\bm\omega}_{Th}^{(\mu)}$ that comes from the Thomas precession,  and the muon cyclotron frequency ${\bm \omega}_c^{(\mu)}$.
The quantity of interest is the angular precession frequency $\Omega_{\mu}$ given by
\begin{equation}\label{Omega_e0}
  {\bm \Omega}_\mu  \equiv {\bm\omega}_{Th}^{(\mu)}+{\bm\omega}_{g_\mu}-{\bm \omega}_c^{(\mu)}\,.
\end{equation}
This quantity can be calculated using the results of \cite{LPSPLB} applied in this case to a muon rather than an electron.
We obtain
 \begin{equation}
 {\bm\omega}_{Th}^{(\mu)} =
  \frac{e{\bf B}}{m}\frac{1}{\gamma_{\mu|n}\gamma_n} \, I_\mu
 - \frac{Q{\bf B}}{M} \frac{1}{\gamma_n} I_Q\,, \label{thomas-fin-1}
   \end{equation}
where the relativistic factors $ \beta_{n}, \gamma_{n} $ refer to the motion of the nucleus in the Storage Ring and $\beta_{\mu|n}, \gamma_{\mu|n}$ to that of the muon relative to the nucleus. The other definitions are
 \begin{equation}\label{Ie-IQ-def}
 I_\mu\equiv \frac{\displaystyle{(\gamma_{\mu|n}\gamma_n)^2\left({\bm \beta}_n^2+\frac{{\bm \beta}_{\mu|n}^2}{\gamma_n}+2\Pi+\frac{\gamma_n \Pi^2}{\gamma_n+1}-Y\right)}}{(1+\Pi)[\gamma_{\mu|n}\gamma_n(1+\Pi)+1]}\,,
 \end{equation}
 \[
 I_Q\equiv \frac{(\gamma_{\mu|n}\gamma_n)^2
 \displaystyle{ \left[{\bm \beta}_n^2\left(1+\frac{\gamma_n\Pi}{\gamma_n+1}\right)^2-X\right]}}{\gamma_{\mu|n}\gamma_n(1+\Pi)+1}\,,
 \]
 \[
 Y\equiv \frac{{\bm \beta}_{\mu|n}^2[\gamma_n(2-\cos^2\theta)-\sin^2\theta]}{3\gamma_n^2},\,
 X\equiv \frac{{\bm \beta}_{\mu|n}^2{\bm \beta}_{n}^2\sin^2\theta}{3(\gamma_n+1)}\,,
 \]
where $\Pi={\bm
\beta_n}\cdot {\bm \beta_{\mu|n}}=\beta_{n}\beta_{\mu|n}\cos \theta$.
Equation (\ref{thomas-fin-1}) for ${\bm\omega}_{Th}$ reduces to the standard angular frequency of the muon
moving in the storage ring when $Y=0$ (owing to $[({\bf B}\cdot {\bm \beta}_{\mu|n}){\bm \beta}_{\mu|n}]=0$). In fact, in absence of the charged nucleus ($Q=0$), which implies ${\bm\beta}_{\mu|n}=0$, $\gamma_{\mu|n}=1$, hence ${\bm \beta}={\bm\beta}_{n}$, $\gamma_{n}=\gamma$, we re-obtain from (\ref{thomas-fin-1}) the standard result \cite{jackson}
\begin{equation}\label{stres}
{\bm\omega}_{Th,l}^{(\mu)} = \frac{eB}{m}\frac{\gamma-1}{\gamma}\,.
\end{equation}
The same result is also recovered by setting ${\bm \beta}_n=0$, so that ${\bm \beta}={\bm \beta}_{\mu|n}$ and $\gamma=\gamma_{\mu|n}$.

Muons have an intrinsic magnetic momentum given by ${\bm \mu}_\mu=-g_\mu {\mu}_B {\bf s}$, where $g_\mu$ is the $g$-factor,  $\mu_B$ the Bohr magneton and ${\bf s}$ the muon spin. When placed in an external magnetic field ${\bf B}$, muons acquire an additional potential energy given by $-{\bm \mu}_\mu \cdot {\bf B}$.
Following \cite{LPSPLB}, we find
 \be\label{omegage}
{\bm \omega}_{g_\mu}=-\frac{g_\mu e}{2m}\,{\Upsilon}\, {\bf B}\,,
 \ee
 where

\be\label{Upsilon0}
\Upsilon \equiv 1-\frac{\gamma_{\mu|n}^2({\bm \beta}_{\mu|n}\cdot {\hat {\bf u}_3})^2}{\gamma(\gamma+1)}\,.
\ee

The calculation of $g$-factors, based on BS-QED, can be carried out with accuracy even when the
expansion parameter is $Z\alpha\simeq 0.4$. The results agree with available direct measurements \cite{vogel}. In
particular, the BS-QED calculation given in \cite{blundell} includes radiative corrections of order $ \alpha/\pi$ and exact
binding corrections. It yields
 \begin{equation}\label{gb}
g_\mu =2\left[\frac{1+2\sqrt{1-(\alpha
Z)^2}}{3}+\frac{\alpha}{\pi}C^{(2)}(\alpha Z)\right]\,,
\end{equation}
where
 \[
 C^{(2)}(\alpha Z)\simeq \frac{1}{2}+\frac{1}{12}(\alpha Z)^2+\frac{7}{2}(\alpha Z)^4\,.
 \]
 In the limit $Q=0$ we obtain from (\ref{omegage})
 \begin{equation}\label{lim2}
 {\bm \omega}_{g_\mu,l} = -\frac{eBg_{\mu}}{2m}\,.
 \end{equation}
For the cyclotron frequency of the bound muon we obtain
 \begin{equation}\label{omegacfinale}
 {\bm \omega}_c^{(\mu)}=\Big[-\frac{eB}{m}\frac{\beta_n}{\beta}\frac{1-({\bm \beta}_{\mu|n}\cdot {\bf {\hat u}}_1)^2}{\gamma_{\mu|n}\gamma_n(1+\Pi)^2}
 +\frac{QB}{M} \frac{\beta_n}{\beta}\frac{1}{\gamma_n(1+\Pi)}\left(1+\frac{\gamma_n \Pi}{\gamma_n+1}\right)\Big]{\bf {\hat u}_3}
 \end{equation}
which, in the usual limit $Q=0$, yields
\begin{equation}\label{lim3}
{\bm \omega}_{c,l}^{(\mu)}=\frac{eB}{m\gamma}\,,
\end{equation}
and from (\ref{stres}), (\ref{lim2}) and (\ref{lim3}) we get, in the same limit,
\begin{equation}\label{limit}
{\bm\Omega_{\mu,l}}={\bm\omega}_{Th,l}^{(\mu)}+{\bm\omega}_{g_{\mu,l}}-{\bm \omega}_{c,l}^{(\mu)}=-\frac{eB}{m}\left(-\frac{\gamma-1}{\gamma}+\frac{g_{\mu}}{2}-\frac{1}{\gamma}\right)\,,
\end{equation}
which coincides with (\ref{omegafin}).
By comparing (\ref{lim3}) with (\ref{omegafin}) we find that ${\bm \omega}_{c,l}^{(\mu)}$ corresponds to SRC. It therefore follows that ${\bm \omega}_c^{(\mu)}$ is the desired {\it generalization
of SRC to rotating compound spin systems}.

The angular frequencies  ${\bm\omega}_{Th}^{(\mu)}$, ${\bm\omega}_{g_{\mu}}$ and ${\bm \omega}_c^{(\mu)}$ yield the final expression
of the angular precession frequency ${\bm \Omega}_\mu$
 \begin{equation}\label{Omega_e}
   {\bm \Omega}_\mu  \equiv {\bm\omega}_{Th}^{(\mu)}+{\bm\omega}_{g_\mu} -{\bm \omega}_c^{(\mu)}=
 \end{equation}
 \[
  = -\frac{e {\bf B}}{m_\mu}\left(\frac{g_\mu}{2}\Upsilon-\frac{I_\mu}{\gamma_{\mu|n}\gamma_n}-U\right)-\frac{Q{\bf B}}{M}\frac{I_Q+V}{\gamma_n}\,,
 \]
where $\Upsilon$ is defined in (\ref{Upsilon0}) and
 \begin{equation}\label{Upsilon-U-V}
U\equiv \frac{1-({\bm \beta}_{\mu|n}\cdot {\hat {\bf u}_1})^2}{\gamma_{\mu|n}\gamma_n(1+\Pi)^2}\frac{\beta_n}{\beta}\,,
 \end{equation}
 \[
 V=\frac{\beta_n}{\tilde \beta}\frac{1}{(1+\Pi)}\left(1+\frac{\gamma_n \Pi}{\gamma_n+1}\right)  \,.
 \]
Since in our simplified case we are neglecting magnetic field components along
the $x_2$ and $x_3$-axes, we have ${\bf B} = B {\hat {\bf u}}_3$ and can recast (\ref{Omega_e}) in the form
\begin{equation}\label{Omega_e-versor}
    {\bm \Omega}_\mu = \Omega_\mu {\hat {\bf u}}_3\,,
\end{equation}
where
 \begin{equation}\label{Omegaefinal}
    \Omega_\mu \equiv -\frac{eB}{m}\left[\left(\frac{g_\mu}{2}\Upsilon \epsilon_1 -\epsilon_2\frac{I_\mu}{\gamma_{\mu|n}\gamma_n}-U \epsilon_3\right)+\frac{Z}{A} \frac{m}{m_p}\frac{\epsilon_4I_Q+\epsilon_5 V}{\gamma_n}\right]\,,
 \end{equation}
 $m_p\simeq 0.9$GeV is the proton mass and the $\epsilon$'s are tags introduced to distinguish the various contributions. In particular $\epsilon_1$ tags the $g_{\mu}$ contribution, $\epsilon_2$ and
 $\epsilon_4$ those due to $\omega_{(Th)}^{\mu}$, while $\epsilon_3$ and $\epsilon_5$ refer to $\omega_{c}^{(\mu)}$. On carrying out averages over the angles and taking $\gamma_n \sim1.6$, we obtain
\begin{eqnarray}\nonumber
\Omega_\mu &=& \frac{1}{3 T}1068 B \left\{0.65548 \,\frac{Z}{A} \left(\epsilon_5\frac{0.755213}{\sqrt{1-\frac{0.429654}{\gamma _{\mu|n}^2}}}+\epsilon_4\frac{0.1752+1.15226 \gamma _{\mu|n}^2}{1+1.5256 \gamma _{\mu|n}}\right)\right. \\
&+& 8.86788 \left[-\frac{\,\epsilon_3 \left(0.165009+0.330018 \gamma _{\mu|n}^2\right)}{\sqrt{1-\frac{0.429654}{\gamma _{\mu|n}^2}} \gamma _{\mu|n}^3} - \right. \nonumber \\
& & \quad  \left. -\frac{0.65548\, \epsilon_2 \left(-0.841867+2.16932 \gamma _{\mu|n}^2\right)}{\gamma _{\mu|n} \left(1.+1.5256 \gamma _{\mu|n}\right)}\right. + \nonumber\\
& & \quad + \left. \left. \epsilon_1 \left(1+a_{\mu}\right) \left(1-\frac{0.218493 \left(-1+\gamma _{\mu|n}^2\right)}{\gamma _{\mu|n} \left(1+1.5256 \gamma _{\mu|n}\right)}\right)\right]\right\}\,.
\label{Omegamu}
\end{eqnarray}
The $\epsilon_1, \epsilon_2, \epsilon_3$ terms tend to balance one another (Table I). The larger contributions come from $\epsilon_4$ and $\epsilon_5$ and remain dominant so far as $Z/A\sim 0.5$. The last two terms are not present in the original derivation of SRC in which the fermion is not bound. They are entirely due to the presence of the nucleus to which the muon is attached and their contributions, as mentioned above, can be traced back to $\omega_{Th}^{\mu}$ and $\omega_{c}^{(\mu)}$.

Generalizations of SRC can also be obtained for bound bosons following the procedure outlined above.

\section{Summary and conclusions}
\label{concl}

We have extended SRC to rotating compound spin systems. The generalization has been obtained by means of successive Lorentz transformations on the usual assumption that there exists an infinity  of locally inertial observers and that, therefore, the time scale over which the process takes place is small relative to the acceleration time scale of the observer \cite{mash11}.

The Lorentz factor $\gamma_{\mu|n}$ that appears in (\ref{Omegaefinal}) and (\ref{Omegamu}) is the only free parameter of the entire calculation. It cannot, in fact be determined, as in \cite{LPSPLB}, by comparing (\ref{Omegamu}) with experimental data that, as yet, do not exist. We can however give some estimates. Choosing  $\gamma_{\mu|n}$ equal to the Bohr atom value, given by
\begin{equation}\label{bohr}
 \gamma=\sqrt{1+\frac{(Z\alpha)^4}{4}}+\frac{(Z\alpha)^2}{2}\,,
 \end{equation}
we obtain the value $ \gamma_{\mu|n}=1.00011$ that substituted in (\ref{Omegamu}) gives the results reported in Table I.
\begin{table}[t]
\begin{center}
\caption{Partial and total contributions to $\Omega_{\mu}$ and comparison of ${}^4$He$^{1+}$ with a few ions for which $Z/A\sim 1/2$.}
\begin{tabular}{ccccc} \hline\hline
$\epsilon$(Hz) &  ${}^4$He$^{1+}$ &  ${}^{142}$Pr$^{60+}$ & ${}^{140}$Pm$^{58+}$ &  ${}^{122}$I$^{52+}$ \\  \hline\hline
  & & & &  \\
$\epsilon_1 $ & $5.6374\, 10^7$  &  $5.1851\, 10^7$ &$5.1534\, 10^7$ & $5.2735\, 10^7$ \\ \hline
  & & & &  \\
 $\epsilon_2$ &  $-1.9405\, 10^7$ & $-1.9929\, 10^7$ & $-2.0122\, 10^7$ & $-1.9383\, 10^7$  \\ \hline
  & & & &   \\
 $\epsilon_3$ & $-3.6903\, 10^7$ & $-3.1395\, 10^7$ & $-3.0947\, 10^7$ & $-3.2708\, 10^7$  \\ \hline
  & & & &   \\
 $\epsilon_4$ & $1.0940\, 10^6$ & $0.8968\, 10^6$ & $0.9220\, 10^6$ & $0.9066\, 10^6$  \\ \hline
  & & & &   \\
 $\epsilon_5$ & $2.0811\, 10^6$ & $1.7345\, 10^6$ & $1.7605\, 10^6$ &  $1.8105\, 10^6$  \\ \hline \hline
  & & & &   \\
 $\Omega_{\mu}$(Hz) & $3.2417\, 10^6$ & $3.1591\, 10^6$ & $3.1475\, 10^6$ & $3.3613\, 10^6$  \\ \hline \hline
 \end{tabular}
\end{center}
\end{table}
The same value of $ \gamma_{\mu|n}$ leads to the Bohr atom energy $E=-m(1-\gamma_{\mu|n}+(\alpha Z)^2)=-10.8952$KeV for He in good agreement with the relativistic value $E_R=-\frac{m(Z \alpha)^2}{2n^2}(1+\frac{(Z\alpha)^2}{n}(1-\frac{3}{4n}))=-11.2755$KeV. It appears from Table I that a typical modulation frequency $ \sim 3.5\times10^6 $Hz is superimposed on the usual exponential decay of the muon dragged along the ion orbit, while the muon polarization is approximately $\beta \sim \beta_n \sim 0.75$. As a comparison, the typical modulation in the $g-2$ experiment is $\Omega_{\mu}^{(g-2)} \sim 2.3 \times10^5$Hz with a polarization $\beta^{(g-2)} \sim 0.9$.

The difficulties of the experiment, use of muons, incomplete muon polarization, detection of electrons along the Storage Ring, certainly do not suggest that the experiment be tried as an alternative way to measure the anomaly $a_{\mu}\equiv\frac{g_{\mu}-2}{2}$. The problem discussed in this work rather confirms the fact that SRC can generate oscillations in a decay process as in the GSI experiments and underlines the importance that ion accelerators are assuming for fundamental physics. As in the case of the GSI oscillations, SRC links quantum phenomena, like muon decay and $a_{\mu}$, to the
classical parameter $\gamma_{\mu|n}$, which, by itself, deserves attention.







\end{document}